\documentclass[onecollarge,natbib]{svjour2}
\bibpunct{[}{]}{;}{n}{}{,} 
\smartqed  
\usepackage{graphicx}
\journalname{Few-Body Systems (EFB22)}
\begin{document}

\title{One- and two-neutron halos in effective field theory~\thanks{Research 
supported by the US Department of Energy under grant DE-FG02-93ER-40756.}
}


\author{Daniel R. Phillips
}


\institute{Daniel R. Phillips \at
        	      Department of Physics and Astronomy and Institute of Nuclear and Particle Physics, Ohio University, Athens, OH 45701, USA\\
              Tel.: +1-740-593-1698\\
              Fax: +1-740-593-0433\\
                            \email{phillips@phy.ohiou.edu} 
}

\date{Received: date / Accepted: date}

\maketitle

\begin{abstract}
I discuss recent work our group has undertaken on effective-field-theory (EFT) analyses of experimental data pertaining to one- and two-neutron halo nuclei.
\end{abstract}

\section{Introduction}
\label{intro}

``Halo EFT" is built on the scale hierarchy between the typical distances occupied by neutrons in a nuclear halo, $R_{\rm halo}$, and the size of the nuclear core,
$R_{\rm core} \ll R_{\rm halo}$. For nuclei in which the halo consists of a single neutron it is a simple extension of the EFT for shallow $s$-wave bound states developed in, e.g., Refs.~\cite{Ka98,vK99}. However, Halo EFT employs a nuclear core as one of its degrees of freedom, so it typically breaks down at lower energies than does the EFT applicable to shallow few-nucleon bound states. Halo EFT was extended to $p$-wave bound states
in Refs.~\cite{pwaves} and has been applied to various
halo systems including ${}^5$He~\cite{pwaves}, ${}^8$Li~\cite{Rupaketal},
${}^{15}$C~\cite{Rupak:2012},  and a number of two-neutron ($2n$) halos~\cite{CanhamHammer,Rotureau:2013}.

In Ref.~\cite{HP11} we showed Halo EFT is useful for the analysis of data on the Coulomb dissociation 
of one-neutron ($1n$) halos. We used data on the energy levels of ${}^{11}$Be and the B(E1) of its $1/2^+$ to $1/2^-$ transition to fix the leading-order (LO) EFT parameters. We then predicted the Coulomb dissociation spectrum of ${}^{11}$Be. At next-to-leading order (NLO) an additional parameter associated with the asymptotic normalization coefficient (ANC) of the ${}^{11}$Be $1/2^+$ state enters. It can be adjusted to obtain a good description of the low-energy dB(E1)/dE spectrum~\cite{HP11}.
In Sec.~\ref{sec:C19} I describe recent results from a similar treatment of the $s$-wave 1$n$ halo ${}^{19}$C. Fitting the N$^2$LO EFT amplitude for the Coulomb dissociation of ${}^{19}$C to experimental data allows accurate values for the $n{}^{18}{\mathrm C}$ effective-range parameters  to be extracted~\cite{AP13}.

The reaction ${}^7{\rm Li} + n \rightarrow {}^8{\rm Li} + \gamma$ involves capture of a neutron into a $p$-wave halo state. There have been many attempts to describe this process theoretically, mainly driven by its relation to a key process in the chain of solar-neutrino production reactions, ${}^7{\rm Be} + p \rightarrow {}^8{\rm B} + \gamma$. An accurate EFT description of these $A=8$ radiative captures rapidly runs into the difficulty that several inputs are needed already at LO. In Section~\ref{sec:Li8} we describe a method we recently developed to deal with this problem. We employ ANCs obtained in {\it ab initio} calculations as input to Halo EFT. The results agree with data on the capture reaction to within the expected accuracy of a LO calculation in this EFT~\cite{Zhang:2013}.

I turn my attention to two-neutron halos in Sec.~\ref{sec:C22}. A recent measurement by Tanaka {\it et al.} of the rms matter radius of ${}^{22}$C gives: $\langle r_m^2 \rangle^{1/2}=5.4 \pm 0.9$ fm~\cite{Tanaka:2010zza}.
By applying universal relations derived from EFT to $^{22}\mathrm{C}$, and including estimates of higher-order EFT corrections in a treatment where ${}^{20}$C is an inert core, we have put constraints on the poorly-known values of the $^{22}\mathrm{C}$ two-neutron separation energy and $n^{20}\mathrm{C}$ virtual-state energy using this experimental datum~\cite{Acharya:2013B}.

\section{Coulomb dissociation of ${}^{19}$C}
\label{sec:C19}

The $^{18}\mathrm{C}$ ground state has $J^{\pi}=0^{+}$, while the ground state of ${}^{19}$C is now understood to be $1/2^+$ and have a one-neutron separation energy of $\approx 500$ keV, appreciably less than  $S_{1n}({}^{18}{\mathrm C})=4.2$ MeV. ${}^{19}$C is thus a candidate for an $s$-wave neutron-halo state. There is a reasonable separation of scales, with an expected expansion parameter $R_{\rm core}/R_{\rm halo} \approx 0.4$.

\begin{figure*}[h]
\begin{centering}
\includegraphics[width=0.45\textwidth]{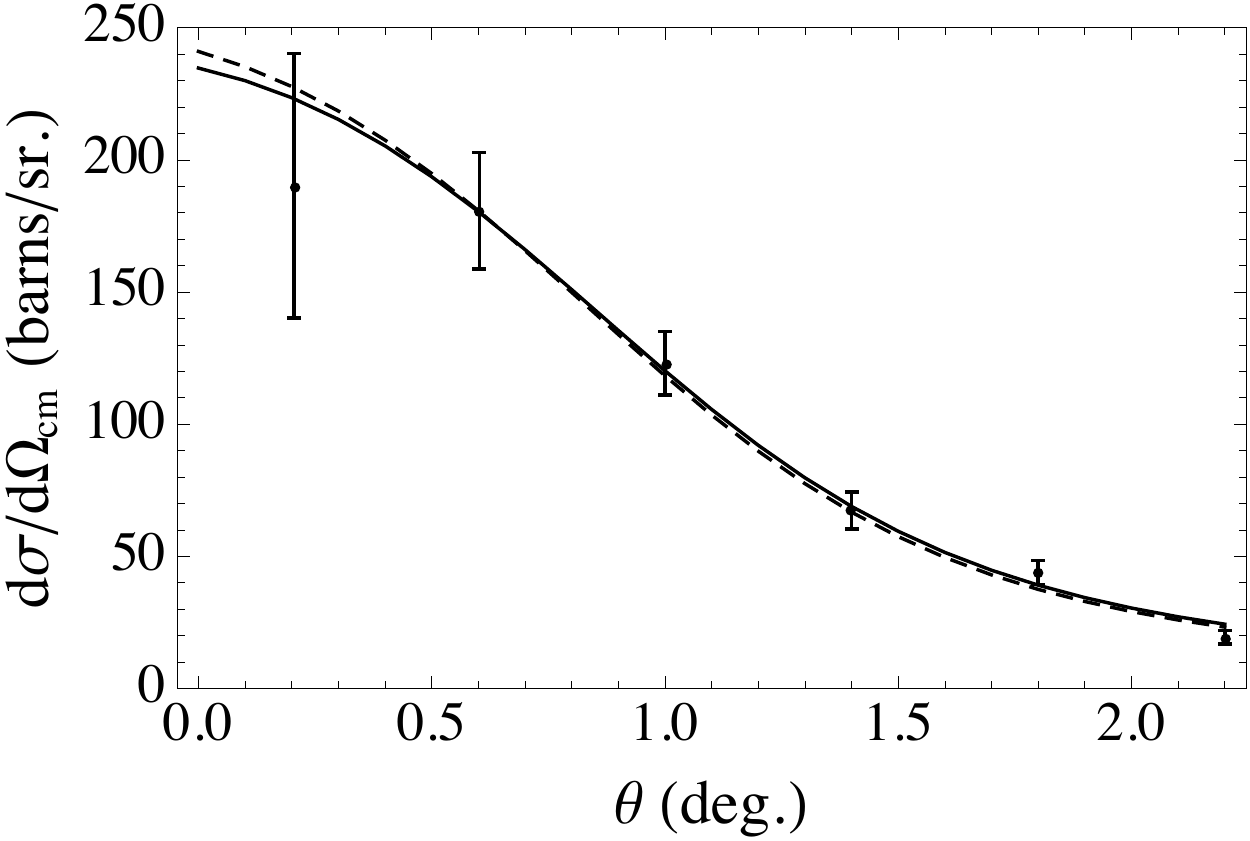}
\includegraphics[width=0.45\textwidth]{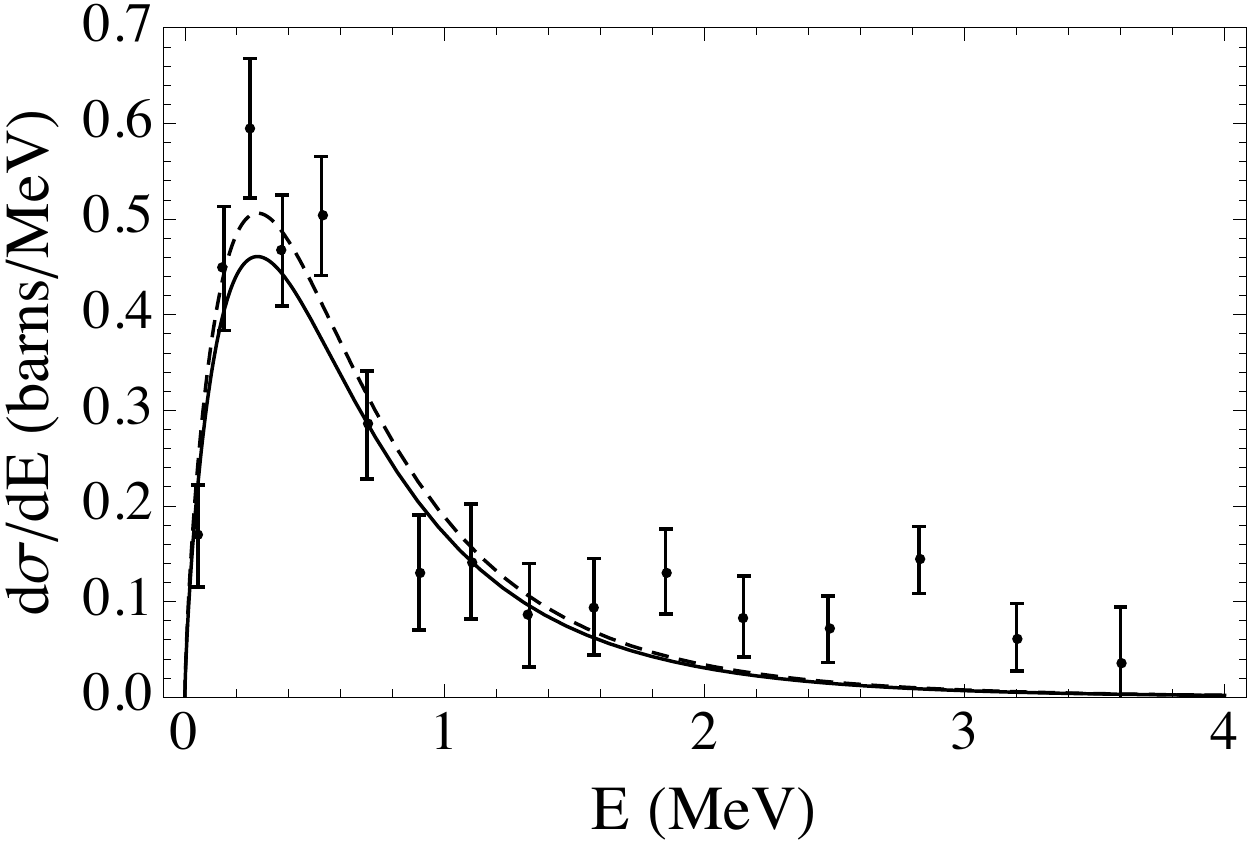}
\caption{(Left panel) Angular distribution of the differential cross section for $S_{1n}({}^{19}\mathrm{C})=540$ keV (dashed), and $S_{1n}({}^{19}\mathrm{C})=575$ keV (solid). Data from Ref.~\cite{Nakamura:1999rp}. (Right panel) Relative energy spectrum of the differential cross section at $S_{1n}({}^{19}\mathrm{C})=580$ keV (dashed), and $S_{1n}({}^{19}\mathrm{C})=575$ keV (solid). Data from Ref.~\cite{Nakamura:2003c}. Figure taken from Ref.~\cite{AP13}.}
\end{centering}
\label{fig:C19}
\end{figure*}

In Ref.~\cite{AP13} we  derived the dipole transition strength, $B(E1)$, for the excitation of $^{19}\mathrm{C}$ to the $^{18}\mathrm{C}+n$ continuum. We found (c.f. Ref.~\cite{Rupak:2012})
\begin{equation}
\label{eq:dbe1overdetoobig}
\frac{\mathrm{d}B(E1)}{e^2 \mathrm{d}E}= \frac{12}{\pi^{2}}\frac{\mu^{3}}{M^{2}}Z^{2} \frac{\gamma_{0}}{1-r_{0}\gamma_{0}}\frac{p^{3}}{\left(\gamma_{0}^{2}+p^{2}\right)^{4}},        
\label{eq:EFTdBE1dE}                                
\end{equation}
where $\gamma_0=\sqrt{2 m_R S_{1n}({}^{19}\mathrm{C})}$ is the binding momentum of the ${}^{18}$C-$n$ bound state, and $r_0$ is the effective range of the $n{}^{18}$C interaction. 
We used Eq.~(\ref{eq:EFTdBE1dE}) as input to a reaction theory that relates dB(E1)/dE to the Coulomb dissociation cross section and fitted the input parameters $\gamma_0$ and $r_0$ to data on the differential angular and differential energy cross sections for Coulomb dissociation of ${}^{19}$C from Refs.~\cite{Nakamura:1999rp} and \cite{Nakamura:2003c}. In Fig.~\ref{fig:C19} we show the input data along with the best fits. The dashed lines in Fig.~\ref{fig:C19} are fits to just the data shown in that panel, and the solid line is the combined fit. The agreement is very good, and extends beyond the fit region $E < 1$ MeV.

We find $S_{1n}({}^{19}\mathrm{C})=(575\pm55(\mathrm{stat.})\pm20(\mathrm{EFT}))$~keV for the one-neutron separation energy and $r_0=(2.6^{+0.6}_{-0.9}(\mathrm{stat.})\pm0.1(\mathrm{EFT}))$~fm. This value of $S_{1n}({}^{19}\mathrm{C})$ is consistent with previous analyses, with a well-defined, small, theory uncertainty. 
The width of the  longitudinal momentum distribution predicted by EFT using these parameters also agrees well with the experimental data of Bazin {\it et al.}~\cite{Bazin:1998zz}. The success of this description affirms the dominance of the $s$-wave configuration of the valence neutron.

\section{Threshold neutron capture on ${}^7$Li}
\label{sec:Li8}

In Ref.~\cite{Zhang:2013} we applied Halo EFT to the $A=8$
capture process ${}^7\mathrm{Li}(n,\gamma){}^8\mathrm{Li}$. The degrees of freedom in our 
calculation are the neutron together with the ground and first-excited states of 
the $^7$Li core.  Halo EFT has
previously been applied to this system
\cite{Rupaketal}, but we used ANCs from\textit{ab initio}
calculations to fix most EFT parameters. The high-energy scale in our EFT is associated with the breakup energy of ${}^7 \mathrm{Li} \rightarrow
t+{}^{4}\mathrm{He}$, $2.5$ MeV.  From the binding energy of ${}^8$Li with respect to the ${}^7$Li-$n$ threshold, $2.03$ MeV, we obtain a nominal expansion parameter
$\sim 0.5$. However, the result we ultimately find for the $p$-wave effective range in $n{}^7$Li scattering suggests a more convergent
expansion. 

We need 7 ANCs to fix the LO parameters of the theory that are pertinent to radiative neutron capture into the ${}^8$Li ground and first-excited states. We take these from Variational Monte Carlo calculations using the Argonne v18 NN potential and the UIX 3N force~\cite{Nollett:2011}. By using the computed ANCs and Halo EFT formulae that relate these to scattering parameters, we find an effective ``range" for $n^7$Li scattering in the channel where ${}^8$Li occurs of
$r_1=-1.43~\mathrm{fm}^{-1}$. 

\begin{figure*}
\begin{centering}
\includegraphics[width=0.6\textwidth]{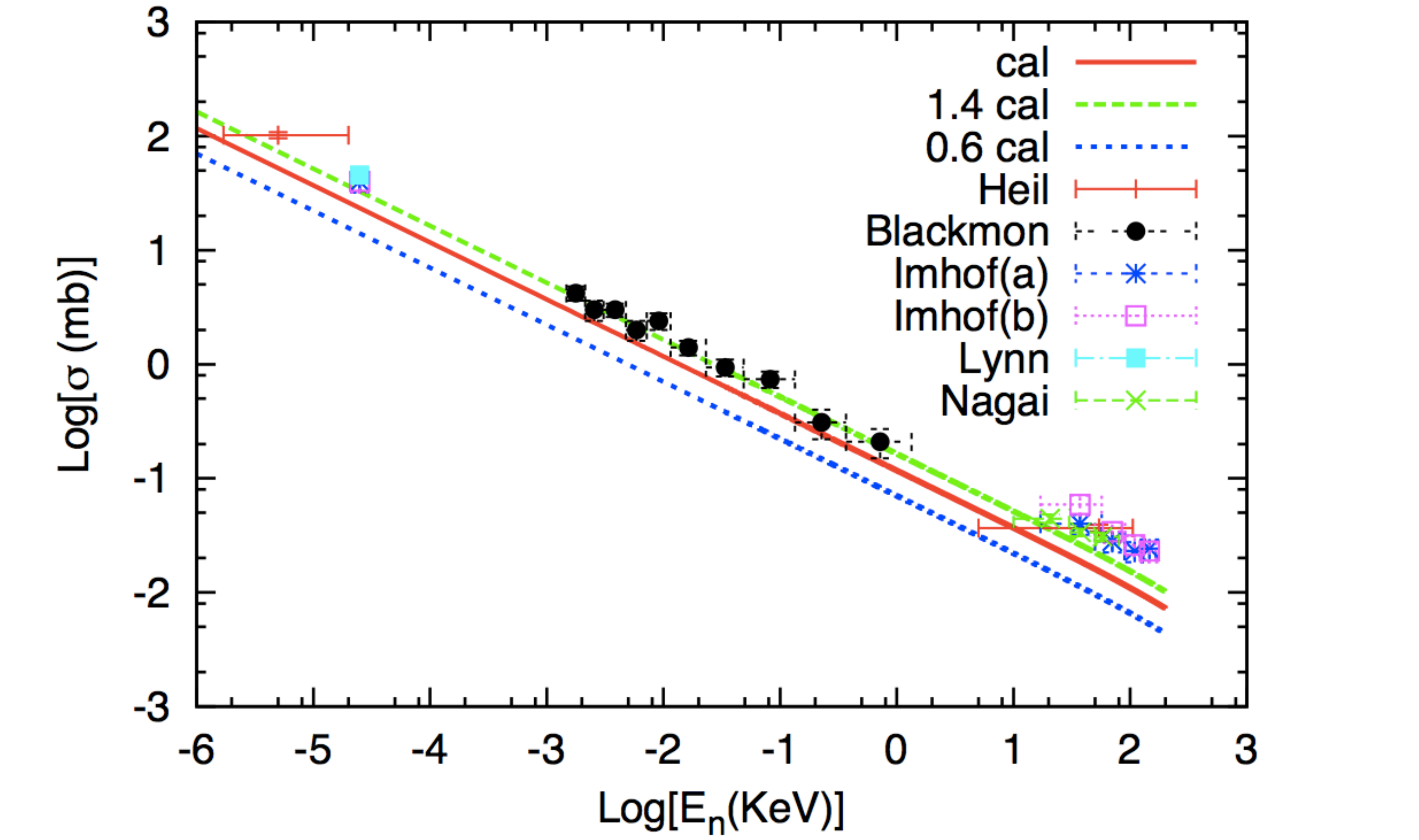}
\caption{Total cross section vs.~neutron lab energy and various data (for details of data see Ref.~\cite{Zhang:2013}). ``cal'' is our LO results, while ``1.4cal'' and ``0.6cal'' are the LO results multiplied by 1.4 and 0.6, to indicate our LO theory uncertainty. Figure taken from Ref.~\cite{Zhang:2013}. } 
\label{fig:xsec}
\end{centering}
\end{figure*}

We can then predict capture to the ground state of ${}^8$Li: results are shown in Fig.~\ref{fig:xsec}. Note that there are two different incoming spin channels: $S_i=2$ and $S_i=1$, and we take account of the large ${}^5$S$_2$ $n^7$Li scattering length by resuming initial-state interactions in that channel. 
The nominal accuracy of our LO cross section is $\approx 40$\%. The central value of our predicted threshold cross section is below the average of the data, but we expect that NLO corrections to the ANCs used in the EFT will push the theory prediction up. Encouragingly, our LO EFT prediction for the ratio of capture into ground and excited states in ${}^8$Li is within 1\% of experiment, and we also obtain a good result for the ratio of capture from different spin channels. This is in contrast to the EFT calculation of Ref.~\cite{Rupaketal}, which made simplifying assumptions about the reaction dynamics. Details will appear in a forthcoming publication~\cite{Zhang:2013}.

\section{The matter radius of ${}^{22}$C}
\label{sec:C22}

\begin{figure}
 \begin{centering}
\includegraphics[width=0.48\textwidth]{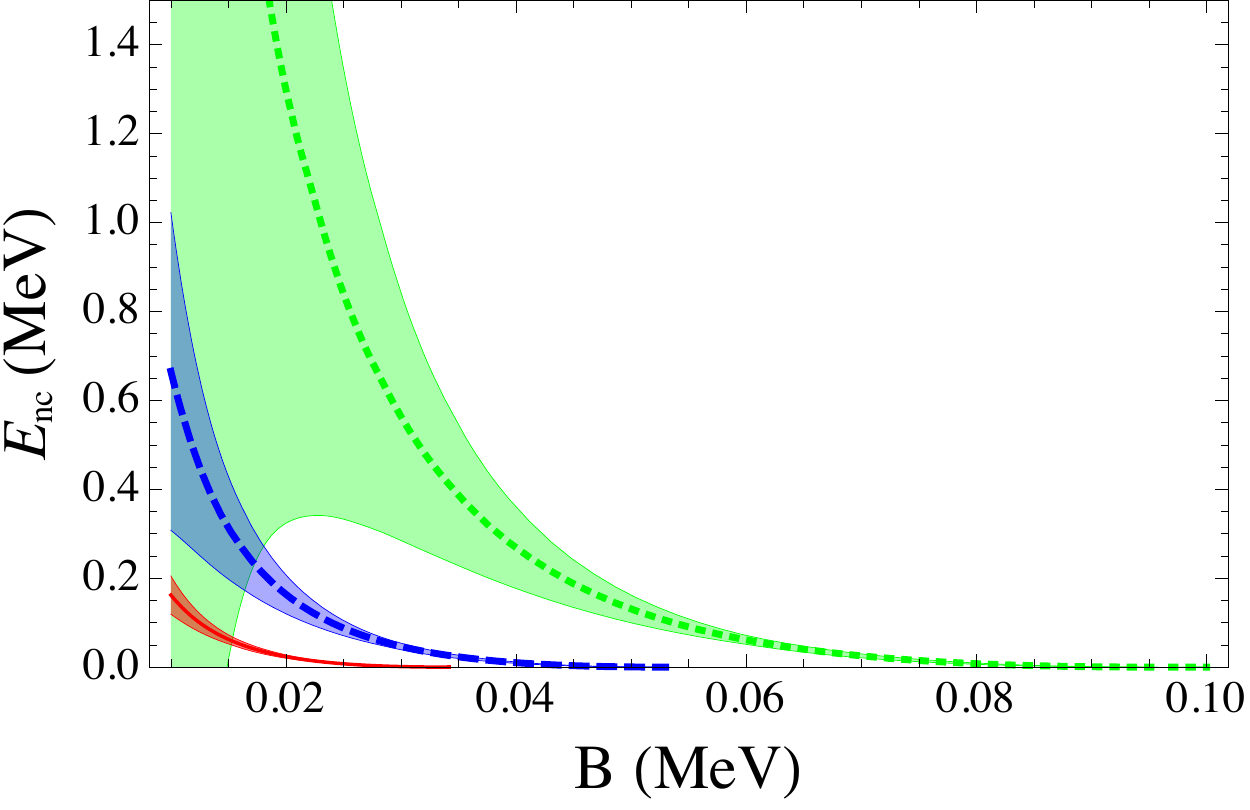}
\caption{Plots of $\sqrt{\langle r^2 \rangle} $~=~5.4~fm (blue, dashed), 6.3~fm (red, solid), and 4.5~fm (green, dotted), with their theoretical error bands, in the $(B,E_{nc})$ plane. Figure taken from Ref.~\cite{Acharya:2013B}.}
\end{centering}
\label{fig:contourplots}
 \end{figure}

EFT was applied to 2{\emph{n}} halo nuclei in Refs.~\cite{CanhamHammer}.
At LO the inputs to the equations describing a 2$n$ halo are the energies of the neutron-core resonance/bound-state, $E_{nc}$, and the $nn$ virtual state, as well as the binding energy, $B$, of the halo nucleus~\footnote{The binding energy of ${}^{22}$C treated as a three-body system is equal to the two-neutron separation energy.}.
In Ref.~\cite{Acharya:2013B} we applied this theory to ${}^{22}$C, understood as a $2n$-halo nucleus with a $^{20}$C core. In Fig.~\ref{fig:contourplots} we plot the sets of  ($B$, $E_{nc}$) values that cover the 1-$\sigma$ range of Tanaka {\it et al.}'s value---$\sqrt{<r^2>}=$ 4.5~fm, 5.4~fm and 6.3~fm---in each case assigning a theoretical error band according to a $R_{\rm core}/R_{\rm halo}$ derived from the matter radius of ${}^{20}$C.  Yamashita {\it et al.} have also investigated the LO correlation between the matter radius and EFT inputs~\cite{Yamashita:2004pv,Yamashita:2011cb}. However, their computation makes additional assumptions about short-distance dynamics and finds a 
matter radius that is too large for a given $B$. Their constraints on  the maximum possible value of $B$ are about 20\% weaker than ours.

Fig.~\ref{fig:contourplots} shows that, regardless of the value of the ${}^{21}$C virtual energy, Tanaka {\it et al.}'s experimental result puts a model-independent upper limit of 100~keV on the 2$n$ separation energy of $^{22}$C. 
The recent experimental finding of Mosby {\it et al.}~\cite{Mosby:2013} that there is no low-energy resonance in the ${}^{21}$C system puts significant tension into this analysis, suggesting that ${}^{22}$C is bound by less than 20 keV (1-$\sigma$, combined EFT and experimental errors). We are presently computing the Coulomb dissociation of ${}^{22}$C in Halo EFT, so that we can predict the structures that will be seen in such data from this system if ${}^{22}$C is indeed this weakly bound. 

\begin{acknowledgements}
I am grateful to my co-authors on the work reported here, Hans-Werner Hammer, Bijaya Acharya, Xilin Zhang, Ken Nollett, and Chen Ji for enjoyable and productive collaborations. I also thank the organizers of EFB22 for an excellently run and very stimulating conference in the beautiful city of Krakow. 
\end{acknowledgements}



\end{document}